\documentclass{article}
\usepackage{spconf,amsmath,amssymb,graphicx,subfigure}
\usepackage{enumitem}
\usepackage{color}
\usepackage[normalem]{ulem}

\title{Bitwise Source Separation on Hashed Spectra: An Efficient Posterior Estimation Scheme Using Partial Rank Order Metrics}
\name{Lijiang Guo,~~~ Minje Kim\thanks{This project was supported by Intel Corporation.}}
\address{Indiana University\\
Department of Intelligent Systems Engineering\\
Bloomington, IN 47408\\ 
\small \texttt{lijguo@indiana.edu}, \texttt{minje@indiana.edu}}
\begin{document}
\ninept
\maketitle
\begin{abstract}
    This paper proposes an efficient bitwise solution to the single-channel source separation task. Most dictionary-based source separation algorithms rely on iterative update rules during the run time, which becomes computationally costly especially when we employ an overcomplete dictionary and sparse encoding that tend to give better separation results. To avoid such cost we propose a bitwise scheme on hashed spectra that leads to an efficient posterior probability calculation. For each source, the algorithm uses a partial rank order metric to extract robust features that form a binarized dictionary of hashed spectra. Then, for a mixture spectrum, its hash code is compared with each source's hashed dictionary in one pass. This simple voting-based dictionary search allows a fast and iteration-free estimation of ratio masking at each bin of a signal spectrogram. We verify that the proposed BitWise Source Separation (BWSS) algorithm produces sensible source separation results for the single-channel speech denoising task, with 6-8 dB mean SDR. To our knowledge, this is the first dictionary based algorithm for this task that is completely iteration-free in both training and testing.
\end{abstract}


\section{Introduction}
\label{sec:intro}
The single-channel source separation problem has been widely studied as a latent variable model. The most common practice is to learn a source-specific dictionary from each source during training so that the source spectra can be reconstructed by a linear combination of the dictionary items. In this way a dictionary defines a discriminative subspace, where its corresponding source spectrum can reside. Using this kind of concept, the source separation procedure for a newly observed mixture spectrum performs another dictionary learning process, where the dictionaries are fixed from the ones the training part, while their activations are estimated using iterative algorithms. Nonnegative Matrix Factorization (NMF) \cite{LeeDD99nature, LeeDD2001nips, RajB2005iwaspaa} and Probabilistic Latent Semantic Indexing (PLSI) \cite{HofmannT99sigir, HofmannT99uai,smaragdis2007supervised} are a popular choice for the modeling job. Meanwhile, a large overcomplete dictionary is another preferable option to preserve the manifold structure of the source spectra. It can be either learned by a manifold preserving quantization technique \cite{kim2013manifold} or simply using the entire source spectra directly as in \cite{smaragdis2009sparse, kim2015efficient}.

As those approaches are based on an iterative algorithm to estimate the activation, a practical source separation system needs to be careful about the necessary resources. Iterative algorithms are not advantageous in two different senses. First, it is not a straightforward decision as to when to stop the iteration unless we have a dedicated predictor for this job \cite{GermainF2015ieeespl}. Second, when it comes to the large overcomplete dictionaries, the accordingly enlarged activation matrix calles for even more computation. 

Deep learning-based solutions tend to predict the separation results in an iteration-free manner by simply running a feedforward pass \cite{HuangP2015ieeeacmaslp,WeningerF2015lvaica,LeRouxJ2015icassp,XuY2014ieeespl,WangY2013ieeeaslp}. However, the single feedforward pass during the test time also needs a lot of resources, e.g. millions of floating-point operations, due to its enlarged structure. Therefore, an efficient dictionary-based solution is still an option especially for a smaller separation problem with a lesser amount of training data.



To this end, a hashing-based speed-up was proposed in \cite{kim2015efficient}, which employes Winner Take All (WTA) hashing \cite{dean2013fast,yagnik2011power} to expedite the reformulated EM updates. It first finds out the nearest neighbors of the current source estimation in the dictionaries based on the Hamming distance between the hashed spectra. Then, it refines the search results by doing a more exact search using cross entropy between the normalized spectra. In this way, the EM updates become faster as their operation can skip non-neighbors in the dictionary. However, it still involves the full cross entropy-based matching procedure as well as the EM iterations. 

In this paper we propose a fully BitWise Source Separation (BWSS) scheme, where the dictionary search is done entirely among the hash codes. To this end, we propose to compare each of the partial rank orders for a randomly chosen magnitude Fourier coefficients of the mixture spectrum with the corresponding one from the source dictionaries, hoping that the partial rank orders of a source is preserved in the mixture. It is based on the W-disjoint orthogonality \cite{rickard2002approximate}, which assumes that there exists a dominant source component in a time-frequency bin. It is convenient that WTA hashing approximately encodes this partial rank orders, so that the dictionary search job during the test time boils down to bitwise operations.





\section{Related Work}
\subsection{Dictionary-based Source Separation}
Dictionary-based source separation methods commonly assume source-specific dictionaries, each of which contains a set of spectral templates that can linearly combine the test mixture. For example, for speech denoising we employ $\mathbf{S}$ and $\mathbf{N}$ that respectively contain $T_S$ and $T_N$ $F$-dimensional dictionary items. The dictionary can be learned by latent variable models, such as NMF \cite{smaragdis2007supervised} or PLSI \cite{duan2012online}, but we use the entire magnitude spectra of the training signals as they are as in \cite{smaragdis2009sparse, kim2013manifold, kim2015efficient}. Once we prepare the dictionaries, the separation job during the test time is to compute the posterior probability of the latent variables at the given time-frequency bin of the mixture spectrum $\mathbf{X}_{f,t}$, namely $P(\mathbf{Z}_{f,t}\!=\!z|\mathbf{X}_{f,t}, \mathbf{S}, \mathbf{N})$, where $\mathbf{Z}_{f,t}$ indicates all the dictionary items from both sources, i.e. $z\in\{1,\cdots,T_S,T_S+1,\cdots,T_S+T_N\}$. Note that the indices are conveniently grouped into the speech and noise parts. In the EM formulation, E-step computes the posterior probabilities as follows:
\begin{equation}
  P(\mathbf{Z}_{f,t}=z|\mathbf{X}_{f,t}, \mathbf{W}=[\mathbf{S}, \mathbf{N}])=\frac{\mathbf{W}_{:,z}\mathbf{H}_{z,:}}{\mathbf{W}\mathbf{H}},
\end{equation}
where $\mathbf{H}$ denotes their activation, which we estimated during the M-step, while $\mathbf{W}=[\mathbf{S}, \mathbf{N}]$ is the concatenated dictionaries that stay fixed. For example, if we adapt PLSI, the update rule for $\mathbf{H}$ is
\begin{equation}
  \mathbf{H}=\frac{\sum_f P(\mathbf{Z}_{f,t}=z|\mathbf{X}_{f,t}, \mathbf{W})\mathbf{X}_{f,t}}{\sum_{f,z} P(\mathbf{Z}_{f,t}=z|\mathbf{X}_{f,t}, \mathbf{W})\mathbf{X}_{f,t}}.
\end{equation}

After the convergence, we eventually consolidate the posterior probabilities to compute the new posterior probability over the two sources $P(\mathbf{Y}_{f,t}=y|\mathbf{X}_{f,t}, \mathbf{W})$, where $y$ indicates one of the two sources: $y=\{0,1\}$. For example,
\begin{align}\label{eq:postDict}
\nonumber  P(\mathbf{Y}_{f,t}=0|\mathbf{X}_{f,t}, \mathbf{W})&={\textstyle\sum_{z=1}^{T_S}}P(\mathbf{Z}_{f,t}=z|\mathbf{X}_{f,t}, \mathbf{W}),\\
  P(\mathbf{Y}_{f,t}=1|\mathbf{X}_{f,t}, \mathbf{W})&={\textstyle \sum_{z=T_S+1}^{T_S+T_N}}P(\mathbf{Z}_{f,t}=z|\mathbf{X}_{f,t}, \mathbf{W}),
\end{align}
which will work like a mask to recover the sources.

Although using larger dictionaries can lead to a better separation \cite{smaragdis2009sparse, kim2013manifold, kim2015efficient}, the computational complexity of the EM-based update rules linearly grows as the size of the dictionaries $T_S$ and $T_N$ become larger. In \cite{kim2015efficient}, this issue was addressed by reformulating the estimation procedure of the activation matrix $\mathbf{H}$ as a nearest neighborhood search problem by using WTA hashing, but it is still based on the EM-based iterative algorithm. This paper investigates an iteration-free dictionary-based method that finds the nearest neighbors in a bitwise manner using the partial rank order as hash codes.

\subsection{Winner Take All (WTA) Hashing}
\label{ssec:WTA}
WTA hashing \cite{yagnik2011power} is a partial rank order based hashing algorithm which has been used to reduce a high dimension feature space to a low dimension feature space while partially preserving the topology of the data. Given a $F$ dimensional feature space WTA hashing first proposes a set of $L$ permutations $\Theta$ whose $\ell^{th}$ entry, $\theta_\ell = \{i_1^\ell, \cdots, i_F^\ell\}$, is a random permutation of the original index, $\{1,\cdots,F\}$. For a data point $\textbf{x} = \{x_1,\cdots,x_F\}$, we index it with $\theta_\ell$ and extract the first $K$ dimensions as a random subset of the $F$ features : $\tilde{\textbf{x}}_\ell = \{x_{i_1^\ell}, \cdots, x_{i_K^\ell}\}$. Let $k_\ell\!=\!\arg\max_k \{x_{i_k^\ell}, 1\leq\!k\!\leq K\}$. Then $x_{k_\ell}$ is the winner of all $K$ feature values of $\tilde{\textbf{x}}_\ell$. We repeat this procedure for all the $L$ permutations in $\Theta$, then we have $L$ integers $\{k_1, \cdots, k_L\}$ as the WTA hash code of $\textbf{x}$.

%

The meaning of $k_\ell$ is worth some discussion as it is closely related to our motivation of using it as a feature for computing similarity measure in later steps. Suppose two data points $\textbf{a}$ and $\textbf{b}$ have the same winning dimension $k_\ell$. This implies that in the $\ell^{th}$ permutation $\theta_\ell$, the same dimension wins over the other same $K-1$ dimensions in $\textbf{x}_1$ and $\textbf{x}_2$. In the magnitude spectrum domain, it means there is a salient peak both at $a_{k_\ell}$ and $b_{k_\ell}$. When comparing each $\ell^{th}$ integer hash code $k_\ell$ of $\textbf{a}$ and $\textbf{b}$, we are estimating a binarized cosine similarity where $1$ means two vectors have same dominant dimension and $0$ means otherwise. The more matched permutation tests are, the more similar $\textbf{a}$ and $\textbf{b}$ are. WTA hashing has shown good performance in object detection \cite{dean2013fast} and source separation \cite{kim2015efficient}.

\section{The Proposed Bitwise Source Separation}
\label{sec:algorithm}
\subsection{Voting-based Likelihood Estimation:\\A Fast Dictionary Search in the Hash Code Space}\label{sec:likelihood}
We propose a nonparametric algorithm for estimating the posterior probability of a signal being one of two sources. To this end, we first calculate the likelihood of observing a time-frequency bin given one of the sources, but based on a simple vote-counting method by finding matches between hashed spectra. This algorithm works on two preprocessed dictionaries of clean speech and noise. For a new mixture spectra, the algorithm scans the two dictionaries to generate a mixture distribution of speech and noise, which can be then used to calculate the posterior probability of one of the sources given the time-frequency bin as in \eqref{eq:postHash}.


For example, suppose there is a magnitude spectrogram of a mixture signal $\mathbf{X}\in\mathbb{R}_+^{F\times T}$ (e.g. a mixture of speech and noise). We first use a partial rank order metric as described in section \ref{ssec:WTA} to generate $L$ integer embeddings of each column vector $\textbf{X}_{:,t}$, call each $\mathcal{X}_{\ell,t}$, where $\ell\in\{1,\cdots,L\}$. The same procedure generates nonnegative integer embedding matrices $\mathcal{S}\in\mathbb{Z}_+^{L\times T_S}$ and $\mathcal{N}\in\mathbb{Z}_+^{L\times T_N}$ for the dictionaries, respectively.

For separation, for each element $\mathcal{X}_{\ell,t}$ we scan $\mathcal{S}_{\ell,:}$ and $\mathcal{N}_{\ell,:}$ to count the number of matches with each dictionary in the $\ell^{th}$ permutation sample, call it $\mathsf{S}_{\ell,t}$ and $\mathsf{N}_{\ell,t}$. Recall $\mathcal{X}_{\ell,t}$ is the integer index of the winning element out of $K$ random dimensions: $\{i_1^\ell,\cdots,i_K^\ell\}$ in the $\ell^{th}$ permutation sample of $\textbf{X}_{:,t}$. Combining $\mathcal{X}_{\ell,t}$ and $\theta_\ell$ we are able to track back to the corresponding original frequency bin $j=i_{\mathcal{X}_{\ell,t}}^{\ell}$, the true winner of $\theta_\ell$ for $\textbf{X}_{:,t}$. Thus the total counts of matches for $j^{th}$ frequency bin with each dictionary that are possibly spread in $L$ slots of $\mathcal{S}_{:,t}$ and $\mathcal{N}_{:,t}$ are defined as follows, respectively: $\bar{\mathsf{S}}_{j,t} =  \sum_{\ell} \mathsf{S}_{\ell,t}$ and $\bar{\mathsf{N}}_{j,t} = \sum_{\ell} \mathsf{N}_{\ell,t}.$

The total counts $\bar{\mathsf{S}}_{j,t}$ and $\bar{\mathsf{N}}_{j,t}$ approximate the similarity of $\textbf{X}_{j,t}$ to the two sources, respectively. Therefore, they also approximate the likelihood of observing $\mathbf{X}_{j,t}$ given one of the sources. In $\ell^{th}$ permutation $\textbf{X}_{i_1^\ell:i_K^\ell,t}$, where $\textbf{X}_{j,t}$ has won, it is greater than the rest $K\!-\!1$ frequencies. Because we encode the rank order of only $K<F$ partial dimensions, the same relationship can be likely to be found in one of the source dictionaries more than in the other. 


\begin{figure}[t]
  \centerline{
  \subfigure[]{\includegraphics[scale=.63]{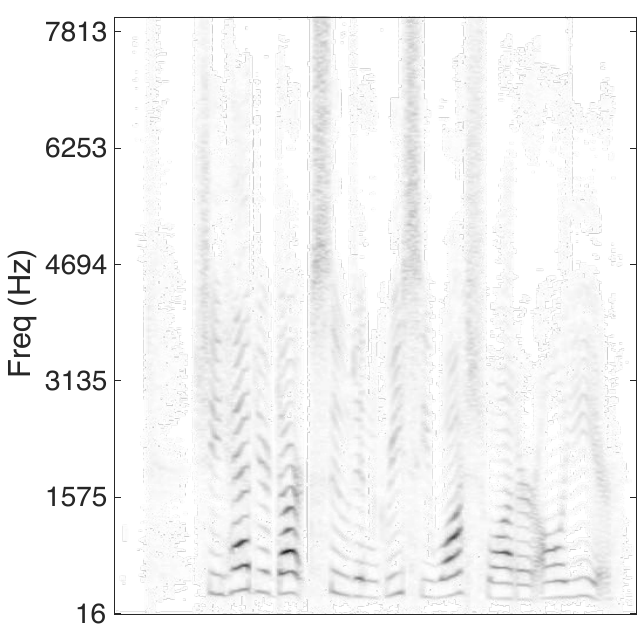}}\hspace{.04in}
  \subfigure[]{\includegraphics[scale=.63]{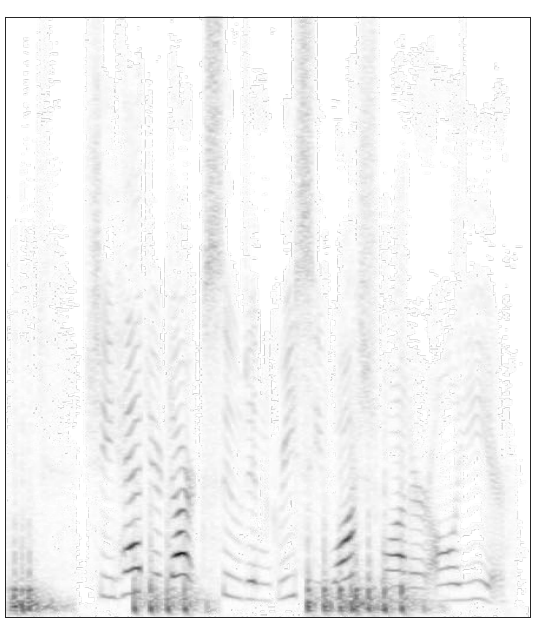}}\hspace{.35in}}
  \centerline{
  \subfigure[]{\includegraphics[scale=.63]{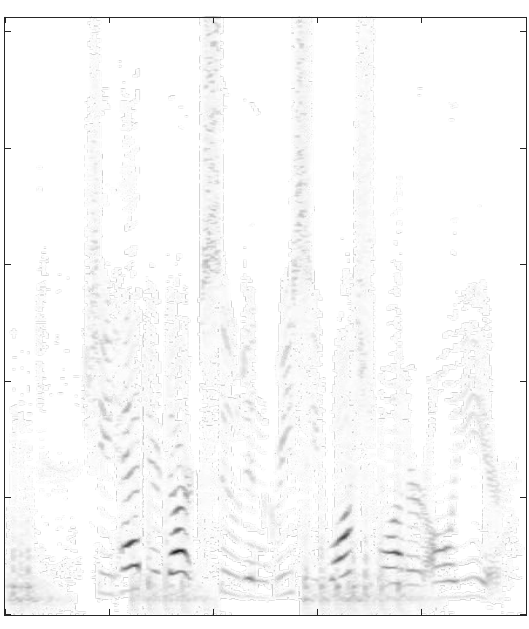}}
  \subfigure[]{\includegraphics[scale=.63]{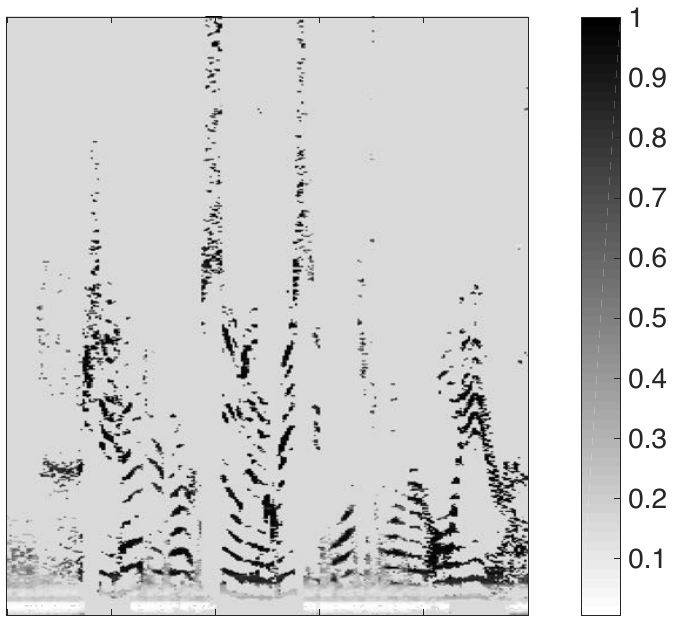}}}
  \caption{Spectrograms and the estimated masks (Female speaker, noise type 8, $K = 64, L = 8$). (a) clean speech (b) noisy speech (c) denoised speech (d) estimated posterior probability mask (1 means speech while 0 is for noise). The original low frequency impulsive noise has been removed.}
  \label{fig:mask}
\end{figure}

\subsection{Estimation of the Posterior Probability}
Once we calculate the likelihoods in the form of the number of partial matches to the two dictionaries as in section~\ref{sec:likelihood}, the rest of the job is to compute the posterior probabilities over the sources given the mixture spectrogram as in \eqref{eq:postDict}. In the proposed BWSS system, we escape from the EM iterations, but instead does the job by calculating the ratio of counts to estimate the posterior probabilities. 

Let $\textbf{Y}_{j,t}$ denote a Bernoulli random variable where $1$ is clean speech and $0$ is noise. Thus the likelihood of observing $\textbf{X}_{j,t}$ is $P(\textbf{X}_{j,t}) = \sum_{\textbf{Y}_{j,t}=\{0,1\}} P(\textbf{Y}_{j,t})P(\textbf{X}_{j,t}|\textbf{Y}_{j,t})$.
We define a prior distribution on $\textbf{Y}_{j,t}$ with a Bernoulli distribution with $p = 0.5$ to give a fair chance to both sources. Another assumption is that each frequency bin is independent of all the other bins in a different time frame, while it is dependent on the other frequency bins in the same time frame due to the rank ordering during hashing. 

To adjust for the difference in the number of frames of the two dictionaries, we normalize the count of matches accordingly. Finally, the posterior probability for a given time-frequency bin is:
\begin{align}
    \label{eq:postHash}
\nonumber&P(\textbf{Y}_{j,t} = 1 | \textbf{X}_{j,t},\mathcal{S}, \mathcal{N}) = 
    \bar{\mathsf{S}}_{j,t}/\big(\bar{\mathsf{S}}_{j,t} + \bar{\mathsf{N}}_{j,t}\cdot r\big),\\
    &P(\textbf{Y}_{j,t} = 0 | \textbf{X}_{j,t},\mathcal{S}, \mathcal{N}) = 
    (\bar{\mathsf{N}}_{j,t}\cdot r)/\big(\bar{\mathsf{S}}_{j,t} + \bar{\mathsf{N}}_{j,t}\cdot r),    
\end{align}
where $r$ is a tuning parameter which depends on the available clean speech and noise used to construct dictionaries; for example, we can set $r = T_S/T_N$. Recall $\bar{\mathsf{S}}_{j,t}$ and $\bar{\mathsf{N}}_{j,t}$ are counts of matches with clean speech and noise dictionaries for a given frequency bin $\textbf{X}_{j,t}$. For $\bar{\mathsf{S}}_{j,t}$, it is the number of votes on the clean speech dictionary for $\textbf{X}_{j,t}$ based on all the permutation samples that $\textbf{X}_{j,t}$ has been involved in comparison and won; similarly $\bar{\mathsf{N}}_{j,t}$ corresponds to the number of votes that $\textbf{X}_{j,t}$ received from the noise dictionary. Thus, $P(\textbf{Y}_{j,t} = 1 | \textbf{X},\mathcal{S}, \mathcal{N})$ reflects the proportion of votes from clean speech dictionary for a frequency bin. Figure~\ref{fig:mask} shows an estimated mask using this posterior probability for source separation.

\subsection{Computational Efficiency}
\label{ssection:comp}
Likewise, we designed an efficient bitwise separation algorithm suitable for resource-efficient environments such as embedded systems. It only requires a single pass per binarized source dictionary for an integer of the mixture hash code, whose complexity is $\mathcal{O}(LKT_ST_N)$. Moreover, the matching can be done using the cheap bitwise AND operation. Therefore, the speed depends on the model parameters $L$ and $K$ as well as the size of the dictionaries.

\section{Experiments}
\label{sec:experiments}

\subsection{The Data Set}
\label{ssec:data}



TIMIT training set contains 136 female and 326 male speakers, while the testing set contains 56 female and 112 male speakers, which are from eight dialect regions in the US. Each TIMIT speaker has 10 short utterances. TSP dataset has over 1400 short utterances from 25 speakers. We downsample the TSP dataset, so that all signals are with a 16kHz sampling rate. We mix each test utterance with 10 kinds of noises as proposed in \cite{duan2012online}. These noises are: 1. birds, 2. casino, 3. cicadas, 4. computer keyboard, 5. eating chips, 6. frogs, 7. jungle, 8. machine guns, 9. motorcycles, and 10. ocean. Short-Time-Fourier-Transform (STFT) with a Hann window of 1024 samples and a hop size of 256 transforms the signals. To evaluate the final results, we used Signal-to-Distortion Ratio (SDR) as an overall source separation measurement along with  Signal-to-Interference Ratio (SIR), and Signal-to-Artifact Ratio (SAR) \cite{vincent2006performance}, and Short-Time Objective Intelligibility (STOI) \cite{taal2010short}.

\subsection{Experiment Design}
\label{ssec:design}
In our experiments, we first construct the hash code dictionaries as described in section~\ref{ssec:WTA}, which yields a set of clean speech dictionaries and 10 noise dictionaries. During source separation, an unseen noisy utterance is processed using the corresponding clean speech dictionary and a noise dictionary of the same noise type. Since the noise type is known, we vary between known and unknown speaker identity to perform supervised and semi-supervised separation. 

Our algorithm has three parameters $K$, $L$, and $r$, and there is no clear guideline in choosing their values. As in dictionary based source separation, \cite{duan2012online} and \cite{sun2013universal} empirically choose the parameters for number of NMF or PLCA basis vectors. For BKL-NMF \cite{sun2013universal}, there is an additional regularization parameter $\lambda$ . We take similar approach to searches for the optimal parameter combination for each testing case. Further details are discussed in section~\ref{ssec:results}.

\textbf{Experiment \#1 Speaker-dependent supervised separation with small dictionaries}: For each TIMIT speaker we use the first 9 out of 10 utterances to create a speaker-specific speech dictionary. By mixing the $10$th one with 10 noise types with 0 dB Signal-to-Noise Ratio (SNR) we get $462\times 10$ noisy test utterances for 462 speakers. Supervised separation was done by assuming the noise type and the identity of the speaker are known. 


\textbf{Experiment \#2 Speaker-dependent supervised separation with large dictionaries}: We suspect that a larger speech dictionary better represents the speaker. For this we use the TSP dataset with roughly 56 utterances per speaker. For each speaker, we once again hold out one last utterance for testing and build the dictionary from the rest. This gives us $25\times 10 = 250$ noisy utterances. Supervised separation is done as in Experiment \#1.

\textbf{Experiment \#3 Pooled-speaker semi-supervised separation}: We apply BWSS in a semi-supervised setting where speaker identity is unknown during separation, while the gender is known. From each dialect of TIMIT training set we select 4 males and 4 females. This gives us $4\times 8 \times 10 = 320$ clean utterances for each gender, which are pooled into one clean speech dictionary for each gender. For testing, we select 1 male and 1 female from each dialect of TIMIT testing set and mix their utterances with 10 noises to create $2\times 8 \times 10\times 10 = 1600$ noisy utterances. During separation, we use the clean speech dictionary of same gender and the noise dictionary of same noise type, which is still a semi-supervised separation with unknown speaker identity. 

\subsection{Separation Results}
\label{ssec:results}

\begin{figure}[t]
  \centering
  \centerline{\includegraphics[width=\columnwidth]{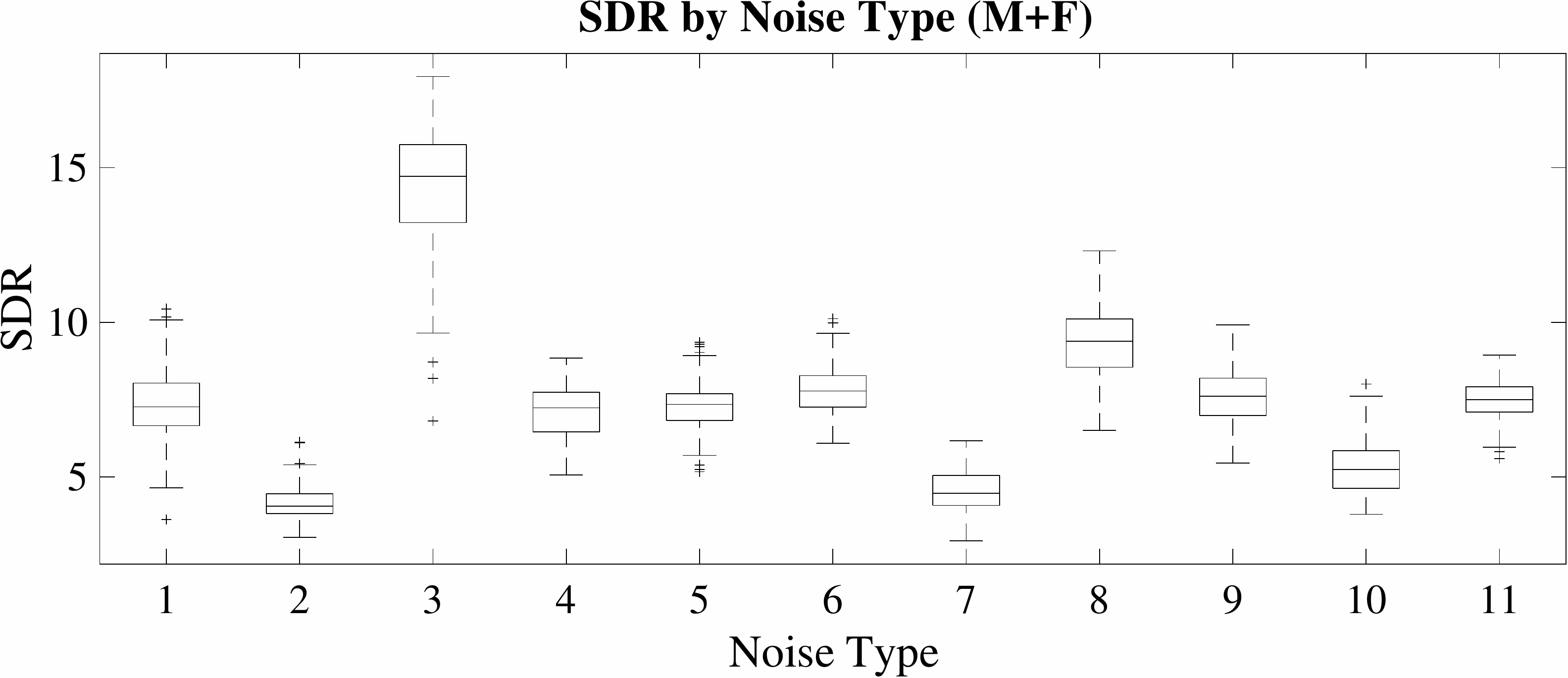}}
  \caption{Mean SDR distributions of 16 speakers in Exp-3. Box 1-10 represent noise type 1-10. Box 11 represents the mean-of-means SDR over 10 noise types.}
  \label{fig:SDR}
\end{figure}

\begin{table}[t]
\centering
\caption{Experiment Results (dB)}
\label{tab:SDR}
\resizebox{0.8\columnwidth}{!}{
 \begin{tabular}{ |c||c|c|c|c| }
 \hline
  & \textbf{SDR} & \textbf{SIR} & \textbf{SAR} & \textbf{STOI}\\
 \hline
 \hline
 \multicolumn{5}{|c|}{\textbf{BWSS Experiment \#1 supervised (TIMIT)}} \\
 \hline
 Male   & 6.6433 & 8.7644 & 9.1727 & 0.0077\\
 Female & 6.7548 & 8.8735 & 9.3551 & 0.0063 \\
 Male and Female    & 6.6761 & 8.7965 & 9.2264 & 0.0073 \\
 \hline
 \hline
 \multicolumn{5}{|c|}{\textbf{BWSS Experiment \#2 supervised (TSP)}} \\
 \hline
 Male and Female    & 6.9898 & 9.3538 & 9.6739 & 0.0128 \\
 \hline 
 \hline
 \multicolumn{5}{|c|}{\textbf{BWSS Experiment \#3 semi-supervised (TIMIT)}} \\
 \hline
 Male   & 7.4213 & 9.7257 & 9.2759 & -0.0104\\
 Female & 7.5271 & 10.343 & 9.4262 & -0.0050\\
 Male and Female    & \textbf{7.4742} & 10.035 & 9.351 & -0.0077\\
 \hline
 \hline
 \multicolumn{5}{|c|}{\textbf{KL-NMF (TIMIT) \cite{sun2013universal}}} \\
 \hline
 Male (supervised)    & 10.23 & - & - & -\\
 Male (semi-supervised)    & \textbf{7.22} & - & - & -\\
 \hline
 \hline
 \multicolumn{5}{|c|}{\textbf{BKL-NMF+USM (TIMIT) \cite{sun2013universal}}} \\
 \hline
 Male (supervised)    & 10.41 & - & - & -\\
 Male (semi-supervised)    & 6.23 & - & - & -\\
 \hline
 \hline
 \multicolumn{5}{|c|}{\textbf{Online PLCA (NOIZEUS) \cite{duan2012online}}} \\
 \hline
 Male and Female    & \textbf{6.180} & 11.710 & 8.450 & -\\
 \hline
\end{tabular}}
\end{table}
\begin{itemize}[leftmargin=0in]
    \setlength{\itemindent}{.15in}
\item {\em \textbf{Variations in parameters}}:
There are three model parameters in the BWSS algorithm, $L$, $K$, and $r$. $L$ is the number of permutation samples to be drawn from a time frame $\textbf{X}_{:,t}$. As $L$ goes to $\infty$, the sample posterior probability will converge to the true mixing distribution. In our experiment we found the algorithm approximates stable posterior probability quickly as we increase $L$. For $L = 2F$ the result is already very close to $L = 8F$. $K$ is the size of each permutation sample. More random samples means the distribution of $\textbf{X}_{:,t}$ is more exploited, and the better approximation to the posterior probability of each frequency bin $\textbf{X}_{j,t}$. However, it is not always guaranteed, because a too large $K$ value can break down the locality of the comparison process. The relative sizes of clean speech dictionary $\mathcal{S}$ and noise dictionary $\mathcal{N}$ are compensated by the parameter $r$. This is because of the possibility that a larger dictionary with more repeating training samples can exaggerate the number of matches for that source. However, note that because of the other chance that the dictionary is indeed with many unique items, the choice of $r$ is not always related to good separation. Also, large $L$ and $K$ increase the computational complexity as discussed in Section~\ref{ssection:comp}. 

To investigate the relation between $L,K,r$ and noise types, we perform grid search for best parameters for different noise types, which was done during Exp. \#3. In there, each \{noise type, gender\} pair has its own optimal $(L,K,r)$, e.g. $(8F,16,0.4)$ for \{noise type 3,  female\}, which is shared across all female test speakers. The effect of noise type on separation performance is shown in Fig. \ref{fig:SDR}. On the other hand, in Exp. \#1 and \#2 the search is for each speaker-noise pair as the speaker identity is assumed to be known.

\item {\em \textbf{Size of clean speech dictionary}}: In a fully supervised setting, both the speaker identity and noise type are known, and the proposed algorithm achieves 6.68 dB mean SDR on TIMIT dataset (Exp. \#1) and 6.99 dB mean SDR on TSP dataset (Exp. \#2), as shown in Table \ref{tab:SDR}. Although in Exp. \#2, each clean speech dictionary has roughly 5 times more speakers than Exp. \#1, we see the performance gap is not very large. Therefore, we conclude that the BWSS algorithm works reasonably well on small, but quality speech dictionaries.

\item{\em \textbf{Speaker identity and pooled dictionary}}: We notice that knowing speaker identity does not provide significant improvement in separation performance compared to the semi-supervised setting with a very large dictionary. For Exp. \#3 the test speaker is unseen, but noise type is known in advance, where the proposed algorithm achieves a mean SDR of 7.47 dB which is 0.80 dB higher than in supervised setting (Exp. \#1). Note that this gender-specific dictionary of 32 speakers is larger than USM's 20 speaker model \cite{sun2013universal}, and would be computationally demanding to handle if it were not for the proposed bitwise mechanism.


\item {\em \textbf{Comparison with other dictionary based methods}}: In Table \ref{tab:SDR}, we include results of two NMF-based methods reported in \cite{sun2013universal} and one PLSI-based method reported in \cite{duan2012online} in addition to BWSS results. All these experiments use the same 10 noise types. In \cite{sun2013universal} 20 male speakers from TIMIT were used as training set, learning 10 basis vectors from each speaker. In \cite{duan2012online} 3 male and 3 female speakers from NOIZEUS formed the training set. The proposed algorithm achieves competitive results using hashed spectra, so that it can employ large training data in a memory-saving manner. Also, its iteration free separation is a plus for the run-time efficiency.

Since the experimental setup is different, a fair comparison is not possible to those existing methods, but we can still gauge the performance of BWSS. For example, a completely supervised model where both the speaker identity and noise type is known (KL-NMF supervised), the usual KL-NMF performs very well (10.23 versus 6.99 dB in Exp. \#2). For the semi-supervised case using KL-NMF, a direct comparison is not possible because it assumes unknown noise, while in Exp. \#3 we assumed anonymity of speakers (7.22 versus 7.47 dB). USM catches up the performance by introducing a larger dictionary and the block sparsity as regularizer, whose supervised case loosely corresponds to Exp. \#3 (10.41 versus 7.47 dB). 

Another comparison would be with \cite{duan2012online}, where an online PLCA algorithm was proposed, but tested with a different speech dataset. For a rough comparison, in all three BWSS experiments, we obtained better SDR and SAR than online PLCA, but marginally worse SIR.

From this comparison, we see that BWSS does not outperform the existing dictionary-based algorithm. Also note that the STOI improvement of BWSS results are not very impressive. However, BWSS performs reasonably well given its low operational cost thanks to its bitwise operations. For example, BWSS can be a viable solution in an extreme environment with little resource. Or, it can be used to better initialize the full NMF/PLCA-based models.




\end{itemize}

\section{Conclusion and Future Work}
\label{sec:conclusion}
We proposed a fully bitwise source separation algorithm. By reformulating the dictionary-based separation algorithm in the binary hash code domain, partial rank orders in particular, we could achieve a nonparametric and iteration-free posterior estimation process which is based on bitwise operations on the binarized feature space. Experiment shows convincing separation results for speech denoising tasks, showcasing the potential of the proposed method in small devices with limited resources. Giving a temporal structure to the algorithm and its application to NMF basis vectors are potentially interesting future directions.

\vfill\pagebreak

\bibliographystyle{IEEEbib}
\bibliography{bwss-refs-17}

\begin{thebibliography}{10}

\bibitem{LeeDD99nature}
Daniel~D Lee and H~Sebastian Seung,
\newblock ``Learning the parts of objects by non-negative matrix
  factorization,''
\newblock {\em Nature}, vol. 401, no. 6755, pp. 788--791, 1999.

\bibitem{LeeDD2001nips}
Daniel~D Lee and H~Sebastian Seung,
\newblock ``Algorithms for non-negative matrix factorization,''
\newblock in {\em Advances in neural information processing systems}, 2001, pp.
  556--562.

\bibitem{RajB2005iwaspaa}
Bhiksha Raj and Paris Smaragdis,
\newblock ``Latent variable decomposition of spectrograms for single channel
  speaker separation,''
\newblock in {\em Applications of Signal Processing to Audio and Acoustics,
  2005. IEEE Workshop on}. IEEE, 2005, pp. 17--20.

\bibitem{HofmannT99sigir}
Thomas Hofmann,
\newblock ``Probabilistic latent semantic indexing,''
\newblock in {\em Proceedings of the 22nd annual international ACM SIGIR
  conference on Research and development in information retrieval}. ACM, 1999,
  pp. 50--57.

\bibitem{HofmannT99uai}
Thomas Hofmann,
\newblock ``Probablistic latent semantic analysis,''
\newblock in {\em Proceedings of the International Conference on Uncertainty in
  Artificial Intelligence (UAI)}, 1999.

\bibitem{smaragdis2007supervised}
Paris Smaragdis, Bhiksha Raj, and Madhusudana Shashanka,
\newblock ``Supervised and semi-supervised separation of sounds from
  single-channel mixtures,''
\newblock {\em Independent Component Analysis and Signal Separation}, pp.
  414--421, 2007.

\bibitem{kim2013manifold}
Minje Kim and Paris Smaragdis,
\newblock ``Manifold preserving hierarchical topic models for quantization and
  approximation,''
\newblock in {\em International Conference on Machine Learning}, 2013, pp.
  1373--1381.

\bibitem{smaragdis2009sparse}
Paris Smaragdis, Madhusudana Shashanka, and Bhiksha Raj,
\newblock ``A sparse non-parametric approach for single channel separation of
  known sounds,''
\newblock in {\em Advances in neural information processing systems}, 2009, pp.
  1705--1713.

\bibitem{kim2015efficient}
Minje Kim, Paris Smaragdis, and Gautham~J Mysore,
\newblock ``Efficient manifold preserving audio source separation using
  locality sensitive hashing,''
\newblock in {\em Acoustics, Speech and Signal Processing (ICASSP), 2015 IEEE
  International Conference on}. IEEE, 2015, pp. 479--483.

\bibitem{GermainF2015ieeespl}
Fran{\c{c}}ois~G Germain and Gautham~J Mysore,
\newblock ``Stopping criteria for non-negative matrix factorization based
  supervised and semi-supervised source separation,''
\newblock {\em IEEE Signal Processing Letters}, vol. 21, no. 10, pp.
  1284--1288, 2014.

\bibitem{HuangP2015ieeeacmaslp}
Po-Sen Huang, Minje Kim, Mark Hasegawa-Johnson, and Paris Smaragdis,
\newblock ``Joint optimization of masks and deep recurrent neural networks for
  monaural source separation,''
\newblock {\em IEEE/ACM Transactions on Audio, Speech and Language Processing
  (TASLP)}, vol. 23, no. 12, pp. 2136--2147, 2015.

\bibitem{WeningerF2015lvaica}
Felix Weninger, Hakan Erdogan, Shinji Watanabe, Emmanuel Vincent, Jonathan {Le
  Roux}, John~R. Hershey, and Bj{\"o}rn Schuller,
\newblock ``Speech enhancement with lstm recurrent neural networks and its
  application to noise-robust asr,''
\newblock in {\em Proceedings of the International Conference on Latent
  Variable Analysis and Signal Separation (LVA/ICA)}, Aug. 2015.

\bibitem{LeRouxJ2015icassp}
Jonathan Le~Roux, John~R Hershey, and Felix Weninger,
\newblock ``Deep nmf for speech separation,''
\newblock in {\em Acoustics, Speech and Signal Processing (ICASSP), 2015 IEEE
  International Conference on}. IEEE, 2015, pp. 66--70.

\bibitem{XuY2014ieeespl}
Yong Xu, Jun Du, Li-Rong Dai, and Chin-Hui Lee,
\newblock ``An experimental study on speech enhancement based on deep neural
  networks,''
\newblock {\em IEEE Signal processing letters}, vol. 21, no. 1, pp. 65--68,
  2014.

\bibitem{WangY2013ieeeaslp}
Yuxuan Wang and DeLiang Wang,
\newblock ``Towards scaling up classification-based speech separation,''
\newblock {\em IEEE Transactions on Audio, Speech, and Language Processing},
  vol. 21, no. 7, pp. 1381--1390, 2013.

\bibitem{dean2013fast}
Thomas Dean, Mark~A Ruzon, Mark Segal, Jonathon Shlens, Sudheendra
  Vijayanarasimhan, and Jay Yagnik,
\newblock ``Fast, accurate detection of 100,000 object classes on a single
  machine,''
\newblock in {\em Proceedings of the IEEE Conference on Computer Vision and
  Pattern Recognition}, 2013, pp. 1814--1821.

\bibitem{yagnik2011power}
Jay Yagnik, Dennis Strelow, David~A Ross, and Ruei-sung Lin,
\newblock ``The power of comparative reasoning,''
\newblock in {\em Computer Vision (ICCV), 2011 IEEE International Conference
  on}. IEEE, 2011, pp. 2431--2438.

\bibitem{rickard2002approximate}
Scott Rickard and Ozgiir Yilmaz,
\newblock ``On the approximate w-disjoint orthogonality of speech,''
\newblock in {\em Acoustics, Speech, and Signal Processing (ICASSP), 2002 IEEE
  International Conference on}. IEEE, 2002, vol.~1, pp. I--529.

\bibitem{duan2012online}
Zhiyao Duan, Gautham~J Mysore, and Paris Smaragdis,
\newblock ``Online plca for real-time semi-supervised source separation,''
\newblock in {\em International Conference on Latent Variable Analysis and
  Signal Separation}. Springer, 2012, pp. 34--41.

\bibitem{vincent2006performance}
Emmanuel Vincent, R{\'e}mi Gribonval, and C{\'e}dric F{\'e}votte,
\newblock ``Performance measurement in blind audio source separation,''
\newblock {\em IEEE transactions on audio, speech, and language processing},
  vol. 14, no. 4, pp. 1462--1469, 2006.

\bibitem{taal2010short}
Cees~H Taal, Richard~C Hendriks, Richard Heusdens, and Jesper Jensen,
\newblock ``A short-time objective intelligibility measure for time-frequency
  weighted noisy speech,''
\newblock in {\em Acoustics Speech and Signal Processing (ICASSP), 2010 IEEE
  International Conference on}. IEEE, 2010, pp. 4214--4217.

\bibitem{sun2013universal}
Dennis~L Sun and Gautham~J Mysore,
\newblock ``Universal speech models for speaker independent single channel
  source separation,''
\newblock in {\em Acoustics, Speech and Signal Processing (ICASSP), 2013 IEEE
  International Conference on}. IEEE, 2013, pp. 141--145.

\end{thebibliography}

\end{document}